# Acoustic Shadow Moiré Interference: Experimental Proof

Ibrahim Abdel-Motaleb[*] and Kiran Vemula
Department of Electrical Engineering, Northern Illinois University, DeKalb, IL 60115
[*]ibrahim@niu.edu

*Abstract-* Shadow moiré techniques are developed for optical waves to image 3D objects to determine their surface topology. In this paper we prove experimentally that also ultrasound can be used to create moiré images. Using moiré techniques will enhance ultrasound applications, such as medical imaging and material characterization. In our experiment, we used a grating mask made of aluminum, an ultrasound source in the megahertz range, and an acousto-optic detector to create and capture Talbot images for the grating. Talbot images are captured using and acousto-optic camera. The captured image was created from ultrasound waves with λ=0.43 mm. The fringes of the images proved that they are shadow moiré fringes.

## I. Introduction

The moiré effect is a mechanical interference of light by superimposed network of lines, or grating [1]. A typical grating structure is shown in Fig. 1, where it has a pattern of broad dark lines separated by transparent lines. The size of the dark lines and the spacing can be in the millimeter to micron range. When light is applied on the gird, a shadow image is overlaid on the surface of the body, as shown in Fig. 2. If the shadow of the grid is observed at the observation point through the same physical grating, a distorted image of the grating is created. This pattern is a moiré pattern formed from the interference between the incident and reflected pattern. Fig. 3 shows an example of an observed moiré pattern, after deleting the noise, [2]. The dark fringe is produced where the dark lines are out of step one-half period, while the bright fringes are produced where the dark lines of one grating fall on top of the corresponding dark lines for the second grating.

The uniqueness of the image is related to the topology of the object surface. This means an algorithm can use the pattern distortion to determine the surface topology of the object. Moiré patterns can also be created if the grating image is observed through a different grating mask with different periodicity, as shown in Fig. 4. The shape of patterns from Fig. 2 and Fig. 4 may be different, but both are related to the surface topology.

Rowe and Welford are the first to report on using shadow moiré fringe pattern to study surface topology [3]. Since that time, shadow moiré has been utilized in many applications. It has been used for studying thermally induced deformation and warpaging in artifacts such as integrated circuits, printed boards, or printed wiring assemblies [4,5,6].

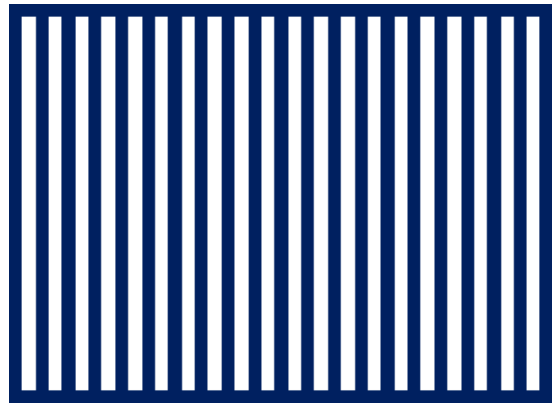

Fig.1. Grating mask that can be used in shadow moiré experiments.

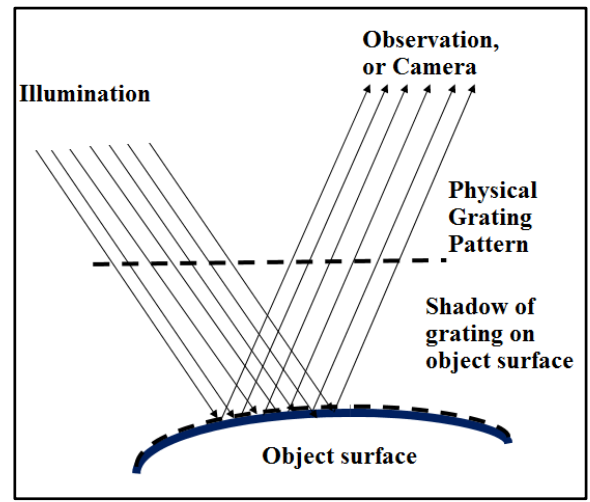

Fig. 2. Shadow of a grid observed through the same grid for the incident light. Observer or camera see a distorted image of grid, or shadow moiré pattern.

Shadow moiré is also used to measure induced stress in microelectronic processes. Chen et al. studied the stress in 500 and 1000 nm $SiO_2$ deposited, using plasma enhanced chemical vapor deposition (PECVD), on top of 525 μm 4-in p-type silicon wafer [7]. The technique used is called Shadow moiré wafer curvature. The experimental results show that this technique can measure warpage within 2 μm of the measurements done using the laser scanning method. The advantage of this technique is that it is based on full-field information, and it would provide a better over results [7].

Moiré topology has found an application in clinical diagnostics [8]. In one application, moiré fringes are projected on the back of a human body. By examining the symmetry of the fringe shadow, abnormalities in the body can be identified [8]. Such evaluation of the image symmetry can be done using software algorithms or by eye observation. This technique is cost effective, noninvasive, fast, and non-radiative that can be easily automated.

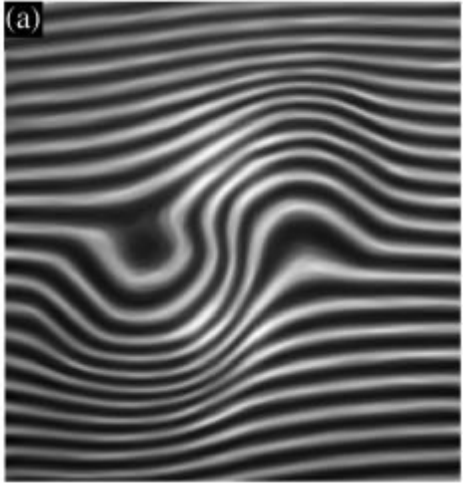

Fig. 3. Noiseless shadow moiré image. Image is used with permission from the IEEE [2].

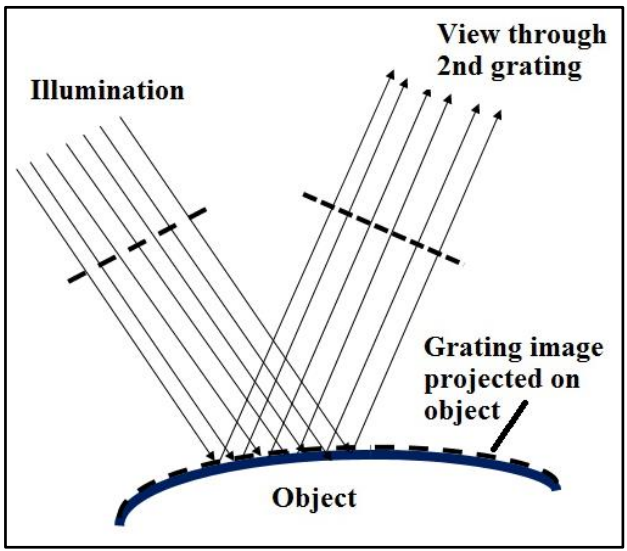

Fig. 4. Set up to observe shadow moire pattern through a second grating pattern.

Using phase-shifting technique, shadow moiré can be employed to obtain the 3D image of an object using a software that utilizes isothermic fringe patterns and digital image processing [9]. A comparison between shadow moiré versus projection moiré shows that the projection moiré can provide about 27 times the accuracy of normal shadow moiré or 27 times pixel density [10].

Employing rigid grating may not provide accurate measurements or inspection for non-flat objects, such as cylinders or cones. Rigid grating can be used effectively with quasi-flat objects, where the difference between the grating and the object is in the range of millimeters. For curved surfaces, flexible grating can provide a more accurate inspection than rigid grating [11, 12]. In this technique a plastic foil with printed Ronchi grating is used with a CCD camera to detect defects, in the range of 30 μm, in curved surfaces.

Optical shadow moiré has many advantages. It is well understood and can provide more resolution than direct imaging. In addition, it provides a quantitative information about the surface topology. However, optical techniques can only investigate the surface for opaque materials. Since many materials are opaque, under surface imaging of defects would be very hard. On the other hand, it is expected that using ultrasound moiré imaging would not have this limitation.

Ultrasound can provide many advantages. For example, ultrasound is safe to use in medical applications. Second, ultrasound can penetrate the body, hence it can be used for inside imaging. Third, acoustic systems are cost effective. Fourth, similar to optical techniques, ultrasound can provide very high resolution, in the micron range, if frequencies around 500 MHz-1 GHz are used.

To realize the above advantage, first, the concept of acoustic shadow moiré should be proven experimentally. This is what we are reporting in this paper. The study objective is to shows that an acoustic moiré images can be created, as a result of the application of an acoustic wave is applied.

II. PRINCIPLES OF SHADOW MOIRÉ

Shadow moiré method is one of the methods in geometric moiré group which can be used to measure out of plane deformations of the surface of an object by giving the contour lines of this 3D object. The grating in front of the object produces a shadow on the object that is viewed from a different direction through the grating itself, as seen in Fig. 5.

The figure shows a linear reference grating of pitch g, placed adjacent to a specimen surface. A light source illuminates the grating and the specimen at an angle α, and a camera receives the light at an angle β. The rays received by the camera are those scattered in its direction by the specimen surface at an angle β. Assuming that the illumination is collimated (light whose rays are nearly parallel) and that the object is viewed through a tele-centric optical system, the height Z between grating and the object point can be determined from the geometry, shown in Fig. 5 [13]. This height is given by,

$$Z = \frac{Ng}{\tan \alpha + \tan \beta} \quad (1)$$

Here α is the illumination angle, β is the viewing angle, g is the spacing of the grating lines and N is the number of grating lines between the points A and B in the figure.

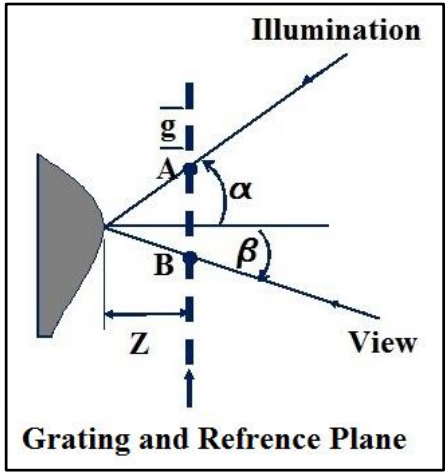

Fig. 5. Geometry for shadow moiré with illumination and viewing at infinity, after [1].

For big objects, it is difficult to illuminate the entire object with a collimated beam. So it is important to consider a case of finite illumination and viewing distances. Fig. 6 shows a geometry where the distance between illumination source and viewing camera is given by W, and the distance between these and the grating is L. The grating is assumed to be close enough to the object surface, so that the diffraction effects are negligible. In this case, the height between the object and the grating is given by

$$Z = \frac{Ng}{\tan \alpha 1 + \tan \beta 1} \quad (2)$$

Here α1 and β1 are the illumination and viewing angles at the object surface. These angles change for every point on the surface and they are different from α and β of Fig. 5.

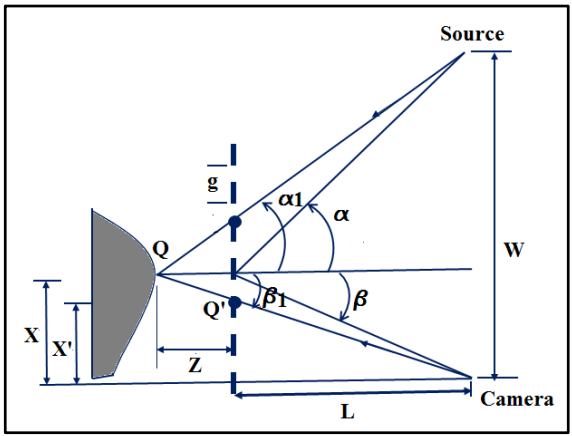

Fig. 6: Geometry for shadow moiré with illumination and viewing at finite distances, after [1].

It should be noted that the governing equation of shadow moiré technique, Eq, (1), is based on the assumption of rectilinear propagation of light, where the light rays travel in straight line, i.e. from a far way source. This assumption is valid only when the gap between the grating and specimen is relatively small compared to Talbot distance [14].

*A. Talbot Planes*

When a plane wave is transmitted through a grating or other periodic structure, the resulting wave front propagates in such a way that it replicates the structure of the grating at multiples of a certain defined distance, known as Talbot length. Talbot was the first to observe this phenomenon, and for this reason the virtual grating was named, Talbot image [15].

Rayleigh derived the mathematical formula expressing the distance between the Talbot images and a linear grating [16]. In this case, normal illumination of a monochromatic plane wave is applied to the grating. For a grating that has a pitch g and is illuminated by a monochromatic collimated light of wavelength λ, the Talbot distance, or self-imaging at normal incidence (α=0°) $D_T^0$, can be obtained from the relation,

$$D_T^0 = \frac{2g^2}{\lambda} \quad (3)$$

In a general case, when the incident light is at angle α, the Talbot distance becomes

$$D_T^\alpha = \frac{2g^2}{\lambda} cos^3 \alpha \quad (4)$$

Talbot images become replica of the grating at Talbot distances of $D_T^\alpha$, $2D_T^\alpha$,.. $nD_T^\alpha$. .., where the image is similar to the grating. On the other hand, at distance of ½$D_T^\alpha$, 1 ½$D_T^\alpha$, ... (n+ ½)$D_T^\alpha$, the images become the complement of the grating, where the dark areas become transparent and the transparent areas become dark. The creation of Talbot images is a result of moiré interference.

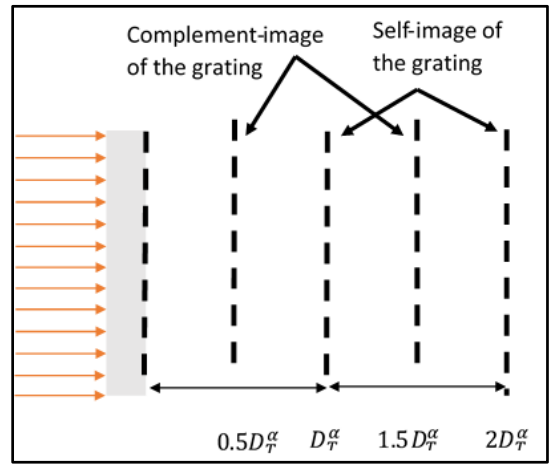

Fig. 7. Talbot images of the grating at ½, 1, 1½, and 2 the Talbot distance. Incident angle α=0.

## III. EXPERIMENTAL RESULTS AND ANALYSIS

To prove that acoustic shadow moiré exists, one can perform one of two experiments. The first is to prove that moiré image is created when ultrasound wave is applied to an object through a grating, as shown in Fig. 6. The second is to show that Talbot images exists, as indicated in Fig. 7. In this paper we report the experimental results of an experiment for creating acoustic Talbot images. This technique is preferred because it is easier and definite.

Talbot images are obtained by placing the grating above the acousto-optic (AO) sensor and below the sound source, as shown in Fig. 8 below. The experiment was performed by setting distance Z to 1 DT and ½ DT. The experiment was performed under normal illumination (α=0° with respect to the normal of the grating). When the Acousto-optic sensor is place at 1 DT from the grating, it captured the image shown in Fig. 9.

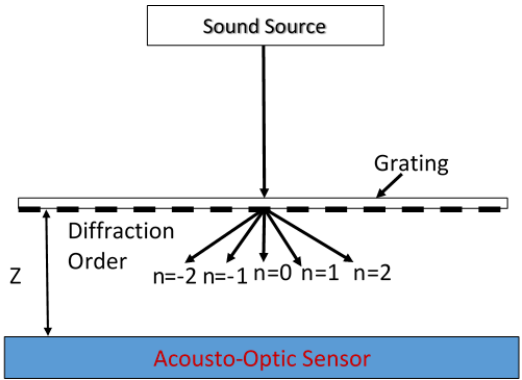

Fig. 8. Set up to investigate the formation of Talbot images.

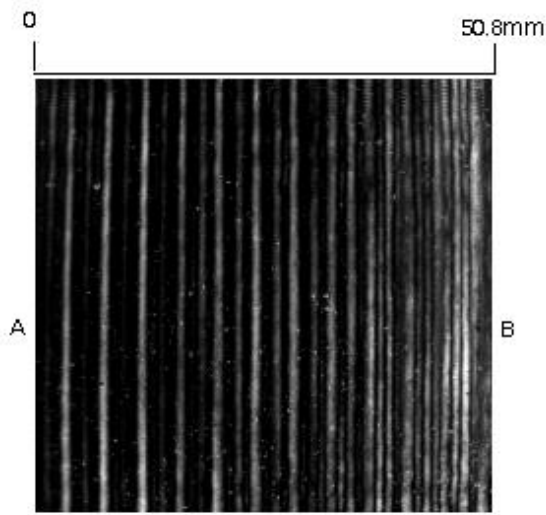

Fig. 9. Diffraction pattern with g=4mm and $f$=3.45 MHz, λ=0.43mm at Z=1 DT (DT=Talbot distance).

From Fig. 9, it can be seen that a grating image has been created. To measure the spacing of the grating lines, we used Matlab to measure the intensity of light of the lines in the image. The Matlab intensity measurement is shown in Fig. 10. From this figure it can be seen that the distance between the grating lines is 4 mm, especially at the Left Hand Side (LHS) of the image. As we move to the Right Hand Side (RHS), lines in between the main lines start to appear. This is an indication that the distance between the grating and the camera deviates from 1DT, as we move to the RHS. The cause of this deviation can be attributed to the fact that the alignment of the grating, the source, and the sensor was done manually. Since grating dimensions and experiment distances are in the millimeter rang, a deviation of few millimeters would be enough to cause distortion. Add to that the possibility that the incident angle changes as we move far from normal line from the source. Inaccurate alignment and change in incident angle becomes more prominent, as we move to the RHS.

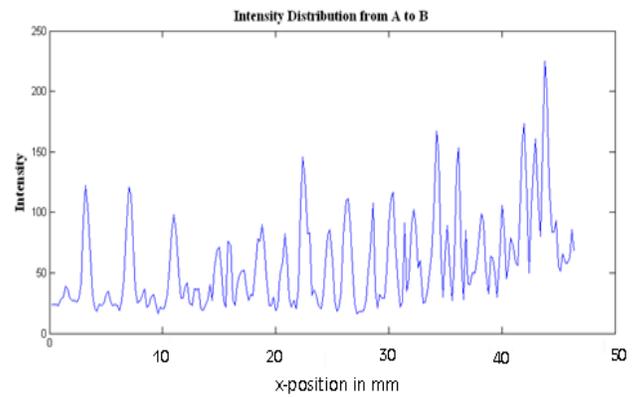

Fig. 10. Intensity distribution along the line from point A to B as shown in Fig. 9.

The sensor was next moved to ½ DT and the image of the grating captured at that distance is shown in Fig. 11. The intensity of the lines of the image was also measured using Matlab. The intensity wave form is shown in Fig. 12.

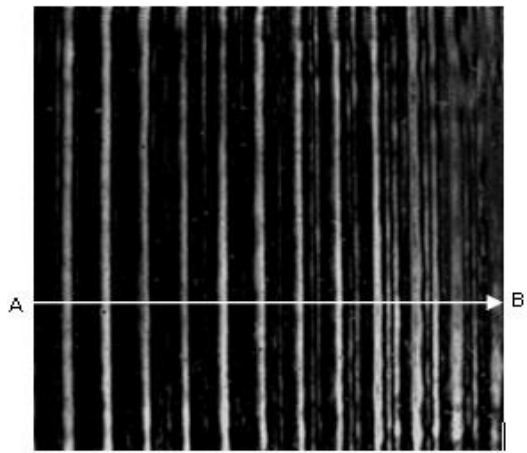

Fig 11. Diffraction pattern at ½ DT, with g=4mm, $f$=3.45 MHz and λ=0.43mm.

Matlab results show that the distance between the fringe lines of the image is 4 mm, which is the same as the physical grating. The images of Fig. 9 and Fig. 11 is expected be identical. If alignment was done accurately, the lines of one of them should be shifted by 2 mm from the other one. Although the two images are identical, as expected, with lines spacing of 4 mm for both, they are not shifted by the expected value of 2 mm. This lack of horizontal shift can be attributed to horizontal misalignment of about 2 mm of the camera, resulting in an image starting 2 mm short.

Similar to Fig. 9, the second image of Fig. 11 shows increased distortion as we move to the RHS. The cause of the distortion is the same reason explained for the first image of Fig. 9. However, the second image is less distorted at the RHS than the first one, and this may be a result of a reduced misalignment in the vertical direction and a better uniformity of the incident angle. Shadow moiré interference has been proven using numerical simulations, as detailed in [17].

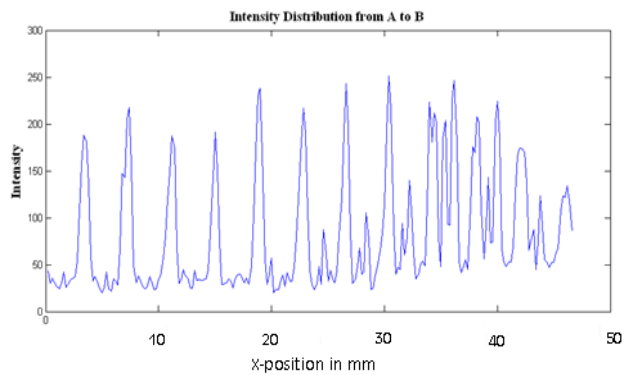

Fig. 12. Intensity distribution along the line from point A to B as shown in Fig. 11.

IV. CONCLUSION

In this study, it has proven experimentally that acoustic shadow moiré interference exits. To increase the resolution to the micrometer instead of the millimeter range, a source with 500 MHz to 1 GHz should be used. Ultrasound has the advantage that it can penetrate almost all materials, including tissues and most industrial materials such as solids. This means, using ultrasound moiré interference would allow for 3D mapping of the inside, as well as the surface of the object. With the low cost and high safety level of ultrasound, it is expected that acoustic moiré interference will have applications in medical diagnosis and industrial inspection.